\documentclass[12pt]{article}
\usepackage{amsmath,amsfonts,amscd}
\usepackage{bm}
\usepackage{graphicx}
\usepackage{lscape}
\usepackage{array,longtable}
\def\ie{{\em i.e.\, }}
\def\etc{{\em etc.\, }}

\def\bra{\langle}
\def\ket{\rangle}

\def\bvec#1{{\bm #1}}

\def\vB{\bvec{B}}

\def\vL{\bvec{L}}

\def\vS{\bvec{S}}

\def\vi{\bvec{i}}
\def\vk{\bvec{k}}

\newcommand{\R}{\mathbb{R}}
\def\Z{\mathbb{Z}}

\def\lr{{L^2(\R)}}

\def\clos{{\rm clos\, }}

\begin{document}
\title{Quantum hierarchic models for information processing}
\author{M.V.Altaisky \\ 
{\small Joint Institute for Nuclear Research, Joliot Curie 6, Dubna, 141980, Russia;} \\
{\small and Space Research Institute RAS, Profsoyuznaya 84/32, Moscow, 117997, Russia}\\
{\small e-mail: altaisky@mx.iki.rssi.ru}
\and 
N.E.Kaputkina \\
{\small National  University of Science and Technology ``MISiS''},\\
{\small Leninsky prospect 4, Moscow, 119049, Russia} \\
{\small e-mail: nataly@misis.ru}
}
\date{Sep 12, 2011}
\maketitle
\begin{abstract}
Both classical and quantum computations operate with the registers of bits.
At nanometer scale the quantum fluctuations at the position of a given bit, say, a quantum dot, not only lead to the decoherence of quantum state of this bit, but also affect the quantum 
states of the neighboring bits, and therefore affect the state of the whole register. 
That is why the requirement of reliable separate access to each bit poses the limit on 
miniaturization, i.e, constrains the memory 
capacity and the speed of computation. In the present paper we suggest an algorithmic  
way to tackle the problem of constructing reliable and compact registers of quantum bits. 
We suggest to access the states of quantum register hierarchically, descending from the state 
of the whole register to the states of its parts. Our method is similar to quantum 
wavelet transform, and can be applied to information compression, quantum memory, quantum 
computations.
\end{abstract}

\section{Introduction}
Classical information can be always encoded in a sequence of bits, the entities 
with two classically distinguishable states. The miniaturization of the 
information processing units to nanometer scales imposes constraints on memory 
capacity and the speed of computation, whatever the algorithm is classical 
or quantum. This constraints arise from the Heisenberg uncertainty principle 
and from the openness of any computational system. The mean momentum transfer required to access an element of 
the size $\Delta x \sim 10^1$nm exceeds $\frac{\hbar}{2\Delta x}$. This corresponds to the electron velocity $v \sim 10^5 {\rm m}/{\rm s}$ and the energy of meV order. The interaction of the same element with the environment at 
room temperature $T \sim 300$K results in energy transfer $k_B T \sim 25$ meV. The decrease of the operating 
voltage to less than one Volt makes nanoscale logical devices to operate with a few electrons only. Thus each 
logical state is achieved not with certainty, but with a finite probability, constrained by quantum effects.

Devoid of the environment a state of quantum bit is linear superposition of two basic states, 0 and 1:
$$
|\psi\ket = \alpha |0\ket + \beta |1\ket,
$$
subjected to a unitary evolution $|\psi(t)\ket = e^{-\imath t H} |\psi(0)\ket$. The interaction with 
environment decoheres the state $|\psi\ket$ into a classical mixture of the two basic states. When the 
interaction with the environment can be neglected the computation can be performed dissipation free by 
unitary evolution of quantum register in a parallel way according to quantum algorithms 
\cite{Nielsen-Chuang:2000,Stolze-Suter:2008}. 
Therefore there exists an 
obstacle that still prevents practical implementation of a workable quantum computer 
with more than a few quantum bits. This obstacle is the 
{\em quantum decoherence} -- the loss of quantum information by means of 
relative dephasing of the qubits in the superposition of quantum states 
due to the interaction with environment. 

The perspective candidates for memory qubits, as well as for quantum gates, are quantum dots -- 
the artificial atoms of $10^1$nm size with the spin of the excess electrons used as quantum bits 
\cite{BLD1999,HKPTV2007}.
The spin of the excess electron in an isolated quantum dot can be reliably initialized 
to a ground state by optical pumping, or by thermal equilibration in a strong magnetic 
field. Having the relaxation time of $10^{-3}$ sec order \cite{KN2000}, the electron spin of a quantum dot,
having the spin decoherence time of $10^{-6}$ sec
order  \cite{KA1998}, can be used as a qubit in quantum computations. 
Challenging problem is to control  spin states of the electrons in an {\em array 
of quantum dots}, separated from each other by less than a typical optical wavelength size, 
rather than an electron spin in a single quantum dot \cite{HKPTV2007}. 

The idea of the present paper is to arrange the quantum bits of a register 
{\em hierarchically} in blocks and process the blocks separately at each 
hierarchy level.  
In the next sections we will develop the necessary formalism to describe 
the Hilbert space of hierarchic states. We also present the physical 
models for hierarchic quantum registers based on the arrays of quantum dots.  

The remainder of this paper is organized as follows. In {\em Section \ref{wt:sec}} we generalize the 
ideas of the Mallat multiresolution analysis to quantum register. In {\em Section \ref{gate:sec}} 
we discuss the CNOT quantum gate in a multiplet basis of quantum register.  In {\em Section \ref{qdot:sec}} 
we discuss physical implementation of quantum gates on spin qubits based on quantum dots. In {\em Conclusion} 
we summarize the advantages of the quantum hierarchic information coding and the difficulties of manipulations 
with such registers.

\section{Generalization of wavelet transform for quantum registers \label{wt:sec}}
Suppose we need to compress a large data vector ($N\!\gg\!1$) in such a way that a few coefficients store the most significant 
information  -- perhaps, that distinguishes the given object from all others, -- 
the next coefficients store some less significant details, \etc Such 
techniques, first proposed by Burt and Adelson   for digital image coding 
\cite{Burt-Adelson:1983}, is known as {\em pyramidal image compression algorithm}.
It is based on the idea, that each four pixels of an image can be considered as 
a block, so that only one value is required to quantify the ``mean color'' of the block, and three more values required to quantify the deviations of pixel colors 
from the block mean. The same procedure can be applied to the group of 4 blocks of 
2$\times$2 pixels each into 16 pixel block, and so fourth. 
The averaging operator 
is usually denoted by $H$, and is referred to as {\em low-pass filter}, the 
projection operator onto the space of averaging-lost details is denoted by 
$G$ and is referred to as {\em high-pass filter}. 
If no information is 
lost during compression, the low- and high-pass filters  obey 
the condition
$
G^* G + H^* H = 1
$. 
$H$ and $G$ operators project the sequence of length $N$ onto the sequences
of length $N/2$ (so, that the total amount of information is conserved), 
decreasing the resolution twice at each step. 
For the one-dimensional data $s \in l^2(\Z)$, the action of $H$ and $G$ 
filters can be written as
\begin{equation}
(Hs)_i = \sum_n h_{n-2i} s_n, \quad (Gs)_i = \sum_n g_{n-2i} s_n.
\label{hg}
\end{equation}
The decomposition of the data vector with the $H$ and $G$ operators 
\eqref{hg} according to the scheme 
$$
\begin{array}{lllllll}
s^0&\stackrel{H}{\to}     & s^1 &\stackrel{H}{\to}     & s^2 & \stackrel{H}{\to}     & \ldots \\
   &\stackrel{G}{\searrow}&     &\stackrel{G}{\searrow} &     &\stackrel{G}{\searrow} &        \\
   &        & d^1 &         & d^2 &         & \ldots
\end{array}, 
$$
and appropriate reconstruction is known as {\em fast wavelet transform  
algorithm} \cite{Daubechies:1988}.

The pyramidal image coding was generalized into the 
Mallat multiresolution analysis (MRA) \cite{Mallat:1986}.
The multiresolution analysis in $\lr$, or the {\em Mallat sequence},  
is an increasing
sequence of closed subspaces $\{ V_j \}_{j\in\Z}, V_j \in \lr$, such that
\begin{enumerate}
\item  $\ldots \subset V_2 \subset V_1 \subset V_0
\subset V_{-1} \subset \ldots $
\item $\displaystyle \clos \cup_{j\in\Z} V_j = \lr$
\item $\displaystyle \cap_{j\in\Z} V_j = \emptyset$
\item The spaces $V_j$ and $V_{j-1}$ are "similar": \\ 
$
 f(x) \in V_j \Leftrightarrow f(2x) \in V_{j-1},\quad j \in \Z$
\end{enumerate}
To set a basis on the Mallat sequence one needs to choose a scaling function
$\phi(x)$, so that
$$V_j = {\rm linear\ span} \{ \phi^j_k; j,k \in \Z \}, $$
where $\phi^0_k(x) \equiv \phi(x-k),$ and 
$\phi^j_k (x) = 2^{-j/2} \phi(2^{-j}x - k)$.
Any function $f\in V_1$, due to the inclusion property 1,
can be written as a linear combination of the basic functions 
of $V_0$.
Since the spaces $V_j$ and $V_{j+1}$ are different in resolution,
some details are being lost when one sequentially projects a function $f \in V_0$
on a ladder of spaces $V_1, V_2, \ldots$. This details can be stored
in the orthogonal complements $W_j = V_{j-1}\setminus V_j$. Explicitly:
$V_0 = W_1 \oplus W_2 \oplus W_3 \oplus \ldots \oplus V_M $.

In quantum case we might also suggest a block structure of application 
of a linear operator to a quantum register. 
However we have first to consider a simpler 
question: How  the qubits can be stored in quantum register, and can 
there exist any quantum structure similar to the Mallat sequence? 
The direct analogs of the Haar wavelet transform, based on quantum networks, 
have been already suggested for quantum computations \cite{hoyer1997,fijany1999quantum},
but they implement a separate access to each quantum bit using the quantum gates, and 
can hardly form an effective memory.

Suppose we have a quantum register consisting of $N=2^M$ qubits and we are going to store 
$N'<N$ quantum bits of information in it. The higher is the ratio $N/N'$, the higher 
fidelity of information storage can be achieved: for more than one qubit can be used to 
store the same information. Such devices may be of practical use if depending 
on real amount of information to be stored different number of quantum bits is allocated.

Let us consider a quantum register implemented 
on spin-half particles:
$$
\otimes_{i=1}^N |s^0_i\ket = \otimes_{i=1}^N (a^0_i |\uparrow\ket^0_i +b^0_i |\downarrow\ket^0_i ),\quad 
|a^0_i|^2+|b^0_i|^2=1. 
$$
In contrast to classical bits the qubits take their values in the $SU(2)$ group,
rather than in $\R$. So, the direct application of the Haar wavelet algorithm 
\begin{equation}
s^j_k = \frac{s^{j-1}_{2k}+s^{j-1}_{2k+1}}{\sqrt 2}, \quad
d^j_k = \frac{s^{j-1}_{2k}-s^{j-1}_{2k+1}}{\sqrt 2}, \quad \label{pd}
k = 0,\ldots,2^{M-j}-1,
\end{equation} 
is not possible. However the spin state 
of a {\em pair of fermions} with spin $\frac{1}{2}$, considered as a compound 
boson, is completely determined by the product $|s_a\ket \otimes |s_b\ket$ 
by means of the Clebsch-Gordan coefficients; and {\sl vice versa}: orthogonality 
of the Clebsch-Gordan coefficients allows to reconstruct the state of the pair of 
fermions from the known state of the boson they comprise 
\cite{Brink-Satchler:1994}.

If a product of $N$ fermion wave functions 
is decomposed into a direct sum of irreducible representations $T^{(J)}$, 
corresponding to the rotations of composite system:
\begin{equation}
\otimes_j D^{(j)}= \oplus_J  c_J T^{(J)}, \label{dsum}
\end{equation}
the compound system may be in either of the states $T^{(J)}$, for which the coefficient $c_J$ is not equal to zero. If a compound system was {\em measured} to be in a state $T^{(m)}$, then only 
those product terms $|s^0\ket \otimes \cdots \otimes |s^0\ket$ can survive, 
which contribute to the $T^{(m)}$ term in \eqref{dsum}.
The process of multiplication of the representations of angular momentum group can be done 
hierarchically, starting from pairs. The whole composite system can be described in 
terms of hierarchic state vectors 
$$
\Phi = \{ |s^M\ket, |s^M s^{M-1}\ket, \ldots ,|s^M s_{M-1}\ldots s^0\ket \}.
$$
In Hilbert space of hierarchic state vectors the reduced density matrix can 
be constructed by taking the trace over the states of the next hierarchy level \cite{Alt03IJQI}.
Hierarchic representation provides an extra possibility to construct quantum gates by 
acting onto the states of the lower hierarchy level $(\alpha\,-\,1)$ 
depending on the states of the next hierarchic level ($\alpha$):
$$
\hat B = |\vi^{(\alpha-1)}\ket |\theta_m^{(\alpha)}\ket B^m_{\vi\vk}\bra\theta_m^{(\alpha)}|  \bra\vk^{(\alpha-1)}|. 
$$
See \cite{Alt03IJQI} for details.

A sequence $\{ s^0_j\}_j$ of bits can be hierarchically compressed 
using the projections onto the ladder of the Mallat sequence 
$V_0,V_{1},\ldots,V_{M}$, where only the projections onto the 
orthogonal complements 
$W_{k} = V_{k-1}\setminus V_{k}$ can be kept in the memory.
For the sequence of $2^M$ quantum bits the information can be encoded 
in the spin state of the whole system 
$
\Psi = \prod_{k=0}^{2^M-1} \psi_k,
$
where $\psi_k$ is a two-component spinor describing the $k$-th qubit. 

Let us consider simplest cases of $M=1$ and $M=2$.

For $\bm{M=1}$ we have a pair of quantum bits. The composite wave function of such 
system 
transforms according to the representation  
\begin{equation}
D_\frac{1}{2} \otimes D_\frac{1}{2} = D_1 \oplus D_0, \label{d2}
\end{equation}
\ie the total system can be either in triplet ($D_1$) or in singlet ($D_0$) 
state, or in their superposition. The bases of the product states 
(l.h.s. of \eqref{d2}) and the composite system states (r.h.s. of \eqref{d2}) 
are related by the linear transform \eqref{aa}.
If the basis in  two-qubit space is chosen as 
\begin{equation}
(e_1,e_2,e_3,e_4) = (|\downarrow\ket |\downarrow\ket,|\downarrow\ket|\uparrow\ket, |\uparrow\ket |\downarrow\ket, |\uparrow\ket |\uparrow\ket), \label{pb}
\end{equation}
and the basis in the space of states of composite boson is chosen as 
\begin{equation}
(s_1,s_2,s_3,s_4) = (|0,0\ket,|1,-1\ket, |1,0\ket, |1,+1\ket),
\label{mb}
\end{equation}
they are related by a linear transform
\begin{equation}
e_i = A_{ik} s_k, \quad
A =
\begin{pmatrix}
0 & 1 & 0 & 0 \cr
-\frac{1}{\sqrt2} & 0 & \frac{1}{\sqrt2} & 0 \cr
 \frac{1}{\sqrt2} & 0 & \frac{1}{\sqrt2} & 0 \cr
0 & 0 & 0 & 1 
\end{pmatrix}. 
\label{aa}
\end{equation}
For this system we can define the spaces $V_0,V_1, W_1$ as follows: 
$
V_0$ is the product space transforming according to 
$D_\frac{1}{2} \otimes D_\frac{1}{2}$; $V_1$ is the triplet state of the composite  
system, which transforms according to $D_1$ representation; $W_1$ is the singlet 
state of the compound system.

For $\bm{M=2}$ the finest resolution space $V_0$ is the span of the four spinor 
product 
$$\Psi = \psi_0 \psi_1 \psi_2 \psi_3,$$ which transforms according to 
$(D_\frac{1}{2} \otimes D_\frac{1}{2} )\otimes (D_\frac{1}{2} \otimes D_\frac{1}{2} )$.

We define $V_1$ as a linear span of the states of maximal spin of each block 
$$
V_1 = D_1 \otimes D_1=  D_2 \oplus D_1 \oplus D_0.$$
In this case the detail space $W_1$ 
is 
$$
W_1 = V_0\setminus V_1 = D_1 \otimes D_0 + D_0 \otimes D_1 + D_0 \otimes D_0.$$
Similarly, the $V_2$ space is the maximal spin state of a next 
level block, which transforms according to 
$D_2$.
The corresponding detailed space is  
$$W_2 = V_1\setminus V_2 =  D_1 \oplus D_0.$$
The total number of degrees of freedom is conserved. 
$
V_0 = W_1 \oplus W_2 \oplus V_2$.
Their dimensions are 
$16 = 7 + 4 + 5$. 

The information encoding in descending order saves the number of operations 
required to set the necessary configuration in $V_0$ space. For instance, in sufficiently strong magnetic field, the state $|2,2\ket \in V_2 $ of the whole system 
uniquely determines configuration of all 4 qubits. The decrease of magnetic 
field results in evolution of qubit pairs into singlet states. 

\section{Gate operations \label{gate:sec}}
The memory on spin states can be exploited in both classical \cite{nomoto1996,orlov1997} and quantum \cite{RSL2000} memory devices.
Classical devices working on Boolean logic with AND and OR operations dissipate an energy 
of at least $k_B T \ln 2$ per logical step. The dissipation imposes a constraint on computation speed.
Quantum computations are time-reversal, they do not dissipate the energy until the read-out of final 
result.

Quantum computation is performed by a sequence of unitary operations applied to the quantum register.
Any conceivable unitary operation in quantum computing can be 
performed by sequential application of single-qubit gates and the two-qubit CNOT 
gate \cite{barenco1995}.  That is why the CNOT quantum gate is a key element of quantum information 
processing. The action of CNOT gate consists in the 
change of the second qubit state, if the first qubit, used as a {\em control}, is 
in the state ``1''. The action of the CNOT gate is defined by the rules, listed in Table ~\ref{t1}. 
\begin{table}
\begin{center}
\begin{tabular}{cccr}
control            & target & resulting & $S_z$\\
qubit              & qubit  & state       &      \\
$|\downarrow\ket$  &$|\downarrow\ket$& $|\downarrow\ket|\downarrow\ket$ & $-1$ \\
$|\downarrow\ket$  &$|\uparrow\ket$  & $|\downarrow\ket|\uparrow\ket$  & $0$ \\  
$|\uparrow\ket$    &$|\downarrow\ket$& $|\uparrow\ket|  \uparrow\ket$& $0$ \\
$|\uparrow\ket$    &$|\uparrow\ket$& $|\uparrow\ket|\downarrow\ket$ & $+1$\\
\end{tabular}
\caption{CNOT gate implemented on two spin qubits. The last column gives the projection 
of the total spin of two qubits to the $z$ axis. $\uparrow$ corresponds to ``1'' (true), and $\downarrow$ corresponds to ``0'' (false). The first qubit in the pair is considered as a {\em control} qubit.  
The value of the second qubit is changed only in case the first qubit 
is in ``true'' state $|\uparrow\ket$; otherwise the state of the second qubit 
is not affected}
\label{t1}
\end{center}
\end{table}

Any two-level quantum system can be used as a quantum bit. Among many suggested implementations of quantum 
bits, {\em viz.} nuclear magnetic resonance \cite{chuang1998experimental}, trapped ions \cite{cirac1995}, cavity electrodynamics \cite{turchette1995}, the usage of the quantum dot electron spin has a number of advantages: the qubit represented by a real $SU(2)$ spin 
is always well defined qubit without a possibility to dissipate; its decoherence time is much longer than 
that of other type qubits: for GaAs the spin decoherence time is of microsecond order 
\cite{KA1998,BLD1999}.
The control over the entanglement in a pair of quantum dot spin qubits can be performed by changing the coupling constant 
$J(t)$ in the Heisenberg Hamiltonian 
\begin{equation}
H_S = J(t) \vS_1 \cdot \vS_2, \label{HH}
\end{equation}
with typical switching time $\tau_s$ of the coupling constant $J(t)$ being of nanosecond order: 
the condition $\int_0^{\tau_s} J(t) dt = J_0 \tau_s = \pi {\ \rm mod\ } 2\pi$ swaps the states of the 
qubits $\vS_1$ and $\vS_2$ \cite{LossDiVincenzo1998PhysRevA.57.120}.
 
To access the states of two spin qubits in the CNOT gate hierarchically, \ie as the states of a composite boson, 
is to use the pair of operators ${\hat S}^2$ and ${\hat S}_z$ of the composite boson.  
In the polarization basis \eqref{pb} the CNOT gate matrix is written in the form 
$$
C = \begin{pmatrix}
1 & 0 & 0 & 0 \cr
0 & 1 & 0 & 0 \cr
0 & 0 & 0 & 1 \cr
0 & 0 & 1 & 0
\end{pmatrix}.
$$
Since the multiplet states $(S,S_z)$ are linearly expressed in terms of the polarization states \eqref{pb}, 
see Eq.\eqref{aa}, the CNOT gate in the multiplet basis \eqref{mb} is given by the matrix 
\begin{equation}
F = A^{-1} C A,\quad F = 
\begin{pmatrix}
\frac{1}{2} & 0 & -\frac{1}{2} & \frac{1}{\sqrt2} \cr
0 & 1 & 0 & 0 \cr
-\frac{1}{2} & 0 & \frac{1}{2} & \frac{1}{\sqrt2} \cr
\frac{1}{\sqrt2} & 0 & \frac{1}{\sqrt 2} & 0 
\end{pmatrix}.
\label{cnot}
\end{equation}
The implementation of the CNOT gate \eqref{cnot} requires mixing between triplet 
and singlet states. This mixing is almost impossible for ordinary atoms, but can be 
easily performed on quantum dots subjected to oscillating electromagnetic field \cite{WMC1992,DH2003,BGDB2008}.

\section{Implementation on quantum dots \label{qdot:sec}}
To build hierarchic memory register (embedded in nanostructure) one needs an array of switching elements, the 
states of which are reliably controlled by external fields. Spin-based devices are promising for such 
applications in both conventional and quantum memory elements \cite{prinz1995,BLD1999,RSL2000}.
The decoherence time of a charge qubit is of nanosecond order, \ie $10^3$ times shorter than that of 
the spin qubit \cite{Huibers1998}. 
On the other hand the desired number of qubits in a quantum dot array can be entangled by changing the 
electromagnetic field acting on the array \cite{HKPTV2007}. Unlike real atoms, the singlet and triplet energy levels of GaAs quantum dots in an array can be easily controlled by changing the magnetic field and the 
interdot distance \cite{kaputkina2010aperiodic,WMC1992,Pfannkuche1993,BKL1996,LK1998a}.

The quantum gates can be implemented either by changing tunneling barrier between neighboring single-electron 
quantum dots \cite{LossDiVincenzo1998PhysRevA.57.120,Waugh1996}, or by monitoring singlet-triplet transitions in two-electron quantum dot by 
means of spectroscopic manipulations \cite{DH2003,BGDB2008,Koppen2009}. Both ways are technologically feasible 
for GaAs heterostructures, where quantum dots with arbitrary number of excess electrons can be formed 
\cite{Tarucha1996,HKPTV2007}. 

\begin{figure}[ht]
\centering \includegraphics[width=6cm]{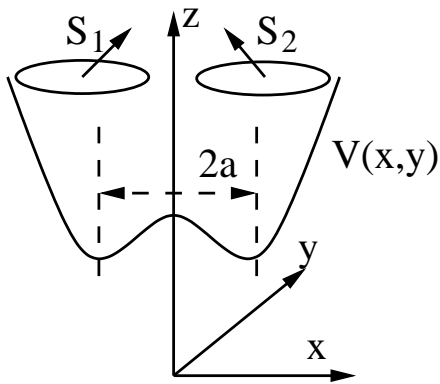}
\caption{Two coupled one-electron quantum dots separated by the distance $2a$ form a quantum gate. Magnetic field $B$ is applied along the $z$ direction. The harmonic wells are centered at $(\pm a,0,0)$. The bias electric field can be applied in $x$ direction}
\label{qd2:pic}
\end{figure}

A realization of quantum gates on the spin degrees of freedom of the coupled quantum dots have been proposed 
in \cite{LossDiVincenzo1998PhysRevA.57.120,BLD1999}. Similarly to the proposed realization of the 
CNOT (CROT) gate on the excitonic excitations in a pair of coupled quantum dots \cite{Li2003Science}, 
a pair of merged quantum dots in a double-well potential, see Fig.~\ref{qd2:pic}, allows for a four 
distinct spin states \eqref{pb}.

The Hamiltonian of coupled single-electron quantum dots has the form 
$$
H = H_{kinetic} + H_{potential} + H_{Zeeman} + H_S,
$$
where $H_{kinetic}$ is the kinetic term, $$H_{potential} = V(x,y)+\frac{e^2}{\varepsilon|r_1-r_2|} + e\sum_{i=1}^2 x_i E$$ includes 
the quantum dot confining potential $V(x,y)$,  Coulomb repulsion of the excess electrons, and the action of 
the bias electric field $E$. The Zeeman splitting term is 
$$H_{Zeeman} = g \mu_B \sum_i \vB\cdot \vS_i,$$ and the Heisenberg Hamiltonian $H_S$ is given by \eqref{HH}.
The confining potential is defined as 
$$
V(x,y) = \frac{m\omega_0^2}{2}
\left[\frac{\left(x^2-a^2 \right)^2}{4a^2}
+y^2
\right].
$$
The typical parameters of a quantum dot in GaAs, described in \cite{BLD1999},   are: 
$$ g \approx -0.44, \hbar\omega_0=3 \hbox{meV}, m=0.067m_e, \varepsilon=13.1.$$
The Bohr radius of harmonic confinement with the above listed parameters is 
$$a_B = \sqrt\frac{\hbar}{m\omega_0} \approx 20 \hbox{nm}.$$
The value of the spin-spin coupling constant \eqref{HH} for a pair of coupled single-electron quantum dots is \cite{BLD1999}: 
$$
J = \frac{\hbar\omega_0}{\sinh \left[2d^2 \left(2b-\frac{1}{b} \right)\right]}
\left[ c \sqrt{b} \left\{  
e^{-bd^2}I_0(bd^2)-e^{d^2 \left(b-1/b \right)}I_0\left(d^2 \left(b-1/b \right) \right)
\right\} + \frac{3}{4b}(1+bd^2)
\right],
$$
where 
$ b = \frac{\omega}{\omega_0}=\sqrt{1+\left(\frac{\omega_L}{\omega_0}\right)^2}$ 
is dimensionless magnetic field, $\omega_L = \sqrt{\frac{eB}{2mc}}$ is the Larmor frequency, $I_0$ is the zeroth-order Bessel function.
Varying the magnetic field $B$ in the range 0-2T one can control the value and the sign of the coupling 
$J$ in the range about $\pm 1$meV. The energy difference between the singlet and the triplet states in 
two-electron quantum dots can be found in \cite{WMC1992}.

Quantum XOR, or the CNOT, gate can be obtained by applying a sequence of operations, consisting 
of single qubit rotations and the swapping of two qubits \cite{LossDiVincenzo1998PhysRevA.57.120}:
\begin{equation}
U_{XOR} = e^{\imath \frac{\pi}{2}S_{1z}} e^{-\imath \frac{\pi}{2}S_{2z}} U_{swap}^\frac{1}{2} 
e^{\imath\pi S_{1z}}  U_{swap}^\frac{1}{2}. \label{swap}
\end{equation}
The swapping of two spin states is provided by the Heisenberg Hamiltonian \eqref{HH} with the 
condition for pulse duration $\int_0^{\tau_s} J(t) dt = J_0 \tau_s = \pi\ {\rm mod\ } 2\pi$ provided swapping the states of the 
qubits $\vS_1$ and $\vS_2$ by unitary operator 
$$
U_{s}(t) = {\rm T}e^{\imath \int_0^t H_S(\tau)d\tau}.
$$
Since the Hamiltonian \eqref{HH} can be expressed in terms of the total spin $\hat S = \hat S_1 +  \hat S_2$:
$$H_S = \frac{J}{2} \left[\hat{S}^2-\frac{3}{2} \right],$$ the coupling constant $J$ determines the energy difference between the singlet, and the triplet state and the swapping operation is performed by action on the spin states 
of the compound two electron system. The read-out of the final state can be performed either by spin-to-charge conversion for a single electron tunneling off the dot \cite{LossDiVincenzo1998PhysRevA.57.120,Engel2004}, or by optical spectroscopy of the quantum dot state \cite{BGDB2008}. The $\sqrt{{\rm swap}}$ operations have been performed experimentally on 
quantum dots with the operation time of 180ps \cite{Petta2005}.

The fluctuations of magnetic field do not affect the coherence of the spin states 
if their length is much greater than the magnetic length of quantum dot, which is of $10^1$ nm order.
We can also neglect the spin-orbital coupling \cite{DVL1998},
$$
H_{SO}=\frac{\omega_0^2}{2mc^2}\vL\cdot\vS,
$$
since $H_{SO}/\hbar\omega_0 \sim 10^{-7}$. As a consequence of this, the dephasing effects caused by charge 
density fluctuations can significantly affect only the charge degrees of freedom, but have little effect on the spin (except for the case when the Coulomb repulsion of the electrons is significant \cite{HuSarma2006}).
The interaction between the spins of different quantum dots  is proportional to the inverse third power of the distance. 
The strength of this interaction can be controlled by making the sequence of quantum dots aperiodic. 
The dipole interaction 
between the spin qubits and the surroundings spins of the environment can be estimated as $(g\mu_B)^2/a_B^3\approx 
10^{-9}$ meV for GaAs quantum dots \cite{BLD1999}, which is very small. The only significant source of dephasing is the hyperfine interaction with nuclear spins, which, however, can be strongly suppressed either by 
dynamically polarizing the nuclear spins, or by applying magnetic field  \cite{BLD1999,SAS2011}. 

\section{Conclusion}
In this paper we present an analog of the Mallat sequence for pyramidal information compression, 
widely implemented in classical computing as {\em discrete wavelet transform} (DWT), for the case of quantum memory on electron spins. The known quantum analogs of the DWT are based on a register of quantum bits connected by 
a quantum network, evaluating sums and differences of the qubit functions, same as for classical Haar wavelet algorithm 
\eqref{pd}. As an any scheme that addresses each qubit separately, such scheme faces the usual scalability problem 
of quantum computation. Our method of information compression uses the multiplet decomposition of the whole 
space of spin states of the memory, instead of addressing each qubit separately. Doing so, we avoid the problem 
of decoherence caused by the local information transmission to a given qubit, which constrains miniaturization 
of information processing devices and implies extra restrictions on geometric equality of memory elements. 

Addressing different spin states of the whole system, rather than different quantum bits, can allow for a 
''flexible'' memory elements (say, on aperiodic sequences of quantum dots \cite{kaputkina2010aperiodic}).
The price paid for such flexibility is the spectroscopic problem to distinguish reliably the spin 
states of quantum system containing 2, 4 and more $2^M$ spins. Since the size of the physical support of 
the group of spins, manipulated spectroscopically, say a group of excess electrons in quantum dots, is comparable to the size of single qubit of the same nature, our method possibly provides a new way of miniaturization 
of memory elements on nanoscale heterostructures. 
  
\section*{Acknowledgment} The authors are thankful to Dr. V.N.Gorbachev for useful discussions.
The research was supported in part by the RFBR Project 11-02-00604-a and the Program of Creation and Development of the National University of Science and Technology ''MISiS''.

\end{document}